\documentclass[11pt, a4paper]{article}
\pdfoutput=1

\usepackage{jcappub}  
\usepackage{amsmath}
\usepackage{amsfonts}
\usepackage{amssymb}
\usepackage{mathrsfs}
\usepackage{graphicx, rotating}
\usepackage{epstopdf}

\usepackage{epsfig}
\usepackage{latexsym}
\usepackage{graphicx}
\usepackage{color}
\usepackage{amsmath,amssymb}
\usepackage{cite}
\usepackage{slashed}
\usepackage{datetime}
\usepackage{hyperref}

\numberwithin{equation}{section}

\hypersetup{colorlinks, citecolor=bluscuro, linkcolor= bluscuro, urlcolor=bluscuro}
\definecolor{rossos}{cmyk}{0,1,1,0.55}
\definecolor{bluscuro}{rgb}{0.15, 0.2, .85}
\definecolor{bluchiaro}{cmyk}{1,.3,0.,0.1}
\definecolor{verdes}{rgb}{0.1, 0.5, 0.1}
\definecolor{myred}{rgb}{0.85, 0, 0}
\definecolor{myblue}{rgb}{0, 0, 0.7}
\definecolor{mygreen}{rgb}{0, 0.45, 0.1}

 \def\be   {\begin{equation}}   \def\ee   {\end{equation}}
 \def\ba   {\begin{array}}      \def\ea   {\end{array}}
 \def\bea  {\begin{eqnarray}}   \def\eea  {\end{eqnarray}}
 \def\bean {\begin{eqnarray*}}  \def\eean {\end{eqnarray*}}
 \def\nn{\nonumber}

\def\tev{\textrm{ TeV}}

\def\lagr{\mathscr{L}}  

\title{{Electroweak bremsstrahlung for wino-like Dark Matter annihilations}}

\author[a]{Paolo Ciafaloni,}
\author[b]{Denis Comelli,}
\author[c,d]{Andrea De Simone,}
\author[e]{Antonio Riotto,}
\author[f]{Alfredo Urbano}

\affiliation[a]{
 Dipartimento di Fisica, Universit\`a di Lecce and INFN - Sezione di
Lecce, \\Via per Arnesano, I-73100 Lecce, Italy}

\affiliation[b]{ 
 INFN,  Sezione di Ferrara, Via Saragat 3, I-44100 Ferrara, Italy}
\affiliation[c]{ 
SISSA, Via Bonomea 265, I-34136 Trieste, Italy
}
\affiliation[d]{ 
INFN, Sezione di Trieste, I-34136 Trieste, Italy
}
\affiliation[e]{
D\'epartement de Physique Th\'eorique and Centre for Astroparticle Physics (CAP),\\
24 quai E. Ansermet, CH-1211 Gen\`eve, Switzerland}

\affiliation[f]{  
Laboratoire de Physique Th\'eorique de l'\'Ecole Normale Sup\'erieure,\\
24 rue Lhomond, F-75231 Paris, France}

\emailAdd{paolo.ciafaloni@le.infn.it}
\emailAdd{comelli@fe.infn.it}
\emailAdd{andrea.desimone@sissa.it}
\emailAdd{antonio.riotto@unige.ch} 
\emailAdd{urbano@lpt.ens.fr}

\abstract{
If the Dark Matter is the neutral Majorana component of a multiplet which is charged under the 
electroweak interactions of the Standard Model, 
its main annihilation channel is into $W^+ W^-$, while the annihilation into light fermions is helicity
suppressed.
As pointed out recently,
the radiation of gauge bosons from the initial state of the annihilation lifts the suppression 
and opens up
an $s$-wave contribution to the cross section.
We perform the full tree-level calculation of Dark Matter annihilations, including electroweak bremsstrahlung,
in the context of an explicit model corresponding to the supersymmetric wino.
We find that the fermion channel can become as important as the di-boson one. This result has
significant implications for the predictions of the fluxes of particles originating from Dark Matter annihilations.
}

\begin{document}

\begin{flushright}
SISSA 01/2012/EP
\end{flushright}

\maketitle

\section{Introduction}

The indirect searches for Dark Matter (DM) are based on detecting excesses in cosmic rays
 produced by annihilations (or decay) of DM in the galactic halo. 
The computation of the cosmic-ray fluxes arriving at Earth can be decomposed into two steps: 
first, one determines the energy spectra of stable SM particles due to the self-annihilations of DM;
then,  these spectra are propagated
from the annihilation region to Earth according to the evolution equations of cosmic rays
(see e.g. Ref.~\citep{PPPC} for more details).
It is clear that the second step is affected by irreducible uncertainties of  astrophysical nature, 
namely the unknown distribution of DM in the halo and the unknown values of the propagation parameters.
However, the dependence on the  particle physics model describing the DM and its interactions with 
the Standard Model (SM) 
only enters into the first step, which is the one we are going to deal with in this paper.

Recently, significant attention has been devoted to include the effects of ElectroWeak (EW) brems\-strahlung
in the calculation of the fluxes from DM annihilations
\citep{paper0, bell1, paper1, bell2, cheung, ibarra1, paper2, iengo, barger, ibarra2,proc} (for earlier studies on the impact of
gauge boson radiation on DM annihilations or cosmic ray physics, see \citep{bergstrom1, bergstrom2, list1}).
This because, contrarily to the naive expectations, the radiation of EW gauge bosons in the
DM annihilation process alters significantly the energy spectra of final particles.
In particular, in the case where the DM is a Majorana fermion, it has been stressed \citep{bergstrom1, paper1, bell1, ibarra1, paper2, ibarra2}
that EW bremsstrahlung has the important effect of lifting the helicity suppression of fermionic final states.

Most of the studies on Majorana DM focus on the DM being a gauge singlet. However, having in mind the weakly
interacting massive particle candidates for DM, for example  the neutralino of the Minimal Supersymmetric Standard Model (MSSM),
it is natural not to restrict oneself to a gauge singlet and consider the possibilty
that the DM is part of a multiplet charged under the EW interactions. Relevant choices are the ones suggested
by the MSSM spectrum:  $SU(2)_L$-triplet (wino-like) and $SU(2)_L$-doublet (higgsino-like).
Concerning the role of EW bremsstrahlung, when the DM belongs to an $SU(2)$ multiplet
even the initial state of the annihilation process can radiate a gauge boson, unlike what happens
in the case of a gauge singlet (bino-like) DM.

In this paper, we consider a wino-like DM candidate and we study the importance of the effect of initial EW radiation
on the self-annihilation cross section and on the cosmic-ray energy spectra.
This  has been already studied in Ref.~\citep{paper2}, using an Effective Field Theory (EFT) approach. 
Although the EFT description of the DM annihilation process allows to grasp the relevant physics, independently of the details of the underlying microscopic model for the interactions, it does not allow 
to study the regime where the DM mass and the cutoff scale of the EFT are very close to each other. 
This corresponds to a situation where  the  sector containing the DM particle and the
mediators of the interactions to the SM extends over a limited mass range.
In this regime 
the EFT  is not reliable anymore, but on the other hand the effects of EW bremsstrahlung become very important,
as we will later see.
To overcome this limitation, we have to resort to a specific model for the DM and its interactions
and we choose the example of  the  MSSM interactions among the wino, the SM fermions and their scalar supersymmetric partners
(a related work for the case of higgsino-like DM has been recently presented in Ref.~\citep{ibarra2};
for wino DM, the inclusion of virtual one-loop and Sommerfeld EW corrections has been studied in Ref.~\citep{iengo}).

In the next section we present the details of the model we use for the calculations, 
while in section \ref{sec:eft} we comment on the validity and the consequences of the EFT approach.
Our main results are discussed in section \ref{sec:complete} and section \ref{sec:conclusions} contains
some concluding remarks.

\section{The model}
\label{sec:model}

Let us introduce the main features of the model we will use throughout the paper.
As already anticipated, we consider the DM particle to be like a pure wino of the MSSM, 
i.e. an electrically neutral Majorana
component in an $SU(2)_L$-triplet $\chi$ with hypercharge $Y_{\chi}=0$.
The interactions with the generic fermion of the SM, 
described by the left handed doublet $L=(f_1, f_2)^{T}$, are mediated by a scalar $SU(2)_L$-doublet 
$\phi$ with hypercharge $Y_{\phi}=1/2$. More explicitly,
\begin{equation}\label{eq:DMfield}
\chi=\left(
       \begin{array}{cc}
         \chi_0/\sqrt{2} & \chi^+ \\
         \chi^- & -\chi_0/\sqrt{2} \\
       \end{array}
     \right),\hspace{1 cm}\phi=\left(
                                   \begin{array}{c}
                                     \phi^+ \\
                                     \phi_0 \\
                                   \end{array}
                                 \right).
\end{equation}
The total Lagrangian of the model is
\begin{equation}\label{eq:L}
\lagr= \lagr_{\rm SM}+ \lagr_{\chi}+ \lagr_{\phi}+ \lagr_{\rm int},
\end{equation}
where to the SM Lagrangian $\lagr_{\rm SM}$ density we added
\begin{eqnarray}\label{eq:pezzilagrangiana}
  \lagr_{\chi} &=& {\rm Tr}\left[\bar{\chi}i\slashed{D}\chi\right]-\frac{1}{2}{\rm Tr}\left[\bar{\chi}M_{\chi}\chi\right], \\
  \lagr_{\phi} &=& \left(D_{\mu}\phi\right)^{\dag}\left(D^{\mu}\phi\right)-M_{\phi}^2\phi^{\dag}\phi,  \\
  \lagr_{\rm int} &=&  -\sqrt{2}y_{\chi}\bar{L}\chi\widetilde{\phi}+{\rm h.c.}=\nonumber\\
 &&  -y_{\chi}\left[
  \bar{f}_1P_R\left(\chi_0\phi_0^*-\sqrt{2}\chi^+\phi^-\right)+\bar{f}_2P_R\left(\chi_0\phi^-+\sqrt{2}\chi^-\phi_0^*\right)
  \right]+{\rm h.c.}\,,
  \label{Lint}
\end{eqnarray}
where $\widetilde{\phi}\equiv i\sigma_2\phi^*$, being $\sigma_{i=1,2,3}$ the usual Pauli matrices. $M_{\chi}$ is the tree level mass of the triplet, 
and we use the convention $P_{R,L}=(1\pm\gamma^5)/2$ for the chiral projectors. $M_{\phi}$ is the tree level mass of the scalar doublet and 
we assume $M_{\phi}\geq M_{\chi}$. 
Moreover, we neglect the mass splitting between the components of the multiplet generated by loop effects \citep{mdm,splitting} which tends to 
make the charged component slightly heavier then the neutral one; the size of this effect is of the order of 100 MeV for a TeV-scale DM mass. 

Despite its simplicity, this toy model provides the subset of the full MSSM lagrangian relevant for DM, 
where $\chi$ is a pure wino and $\phi$ plays the role of a left-handed sfermion (slepton or squark).
The operators in Eq.~(\ref{Lint}) concide with the wino-fermion-sfermion interactions of the MSSM,
provided that one identifies the generic yukawa coupling $y_\chi$ with $g/\sqrt{2}$, where $g$ is the $SU(2)_L$ coupling constant.
The model parameters $M_\chi, M_\phi, y_\chi$ could be related by the relic abundance constraint. However, we do not address the problem 
to compute the relic abundance in this simple model, which is beyond the scope of this paper.
We expect the Sommerfeld effect and the coannihilation channels to play an important role.
In what follows, we will treat the model parameters as being completely independent.

Let us now quickly sketch the behaviour of the  2-body DM annihilation channels allowed by the interactions in Eq.~(\ref{eq:L});
it is useful to expand the annihilation cross section in powers of the relative velocity $v$ (which is $\sim 10^{-3} c$ in our galaxy today)
\begin{equation}\label{velocityexpansion}
v\sigma = a+ bv^2+\mathcal{O}(v^4)\,,
\end{equation}
where $a$ denotes the $s$-wave ($L=0$) contribution, while $b$ the $p$-wave ($L=1$) one. 
For $M_\chi\gg m_W$,  DM predominantly annihilates in $s$-wave into $W^+W^-$
\begin{equation}\label{eq:WW}
v\sigma(\chi_0\chi_0\to W^+W^-)=\frac{g^4}{8\pi M_{\chi}^2}+\mathcal{O}(v^2)\,.
\end{equation}
The other 2-body annihilation channel is into fermions $\chi_0\chi_0\to f\bar f$; by helicity arguments the $s$-wave is proportional to
$(m_f/M_\chi)^2$ and hence very small for light fermions, while the $p$-wave suffers from the $v^2$ suppression.
According to this simple analysis one might naively  conclude that the $W^+W^-$ final state
is the only (or at least the most important) annihilation channel driving  DM phenomenology both in the early Universe and today.
But we will show that initial EW radiation can upset this expectation.

\section{Remarks from an effective field theory point of view}
\label{sec:eft}

Before turning to the results obtained using the model described above, let us pause to make some
comments on the EFT description of DM annihilations with EW radiation
(we refer the reader to Refs.~\citep{paper1,paper2} for more details). 
This will help to highlight the relevance and the scope of the calculations presented in this paper.

If there is a separation of scales  $M_\chi\ll M_\phi$, then it is possible
to integrate out  from the Lagrangian density in Eq.~(\ref{eq:L}) the degrees of freedom of the heavy scalar 
field $\phi$  and end up with effective operators connecting the DM with the SM fields.  Up to dimension-6 operators
we get
\bea
\label{eq:effectiveLagrangian}
\left.\lagr_{{\rm int}}\right|_{\rm dim-6}&=&
-\frac{|y_{\chi}|^2}{2M_{\phi}^2}\left[\sqrt{2}(\bar{f}_2P_R\gamma^{\mu}f_1)
(\bar{\chi}_0\gamma_{\mu}\chi^-) 
-\sqrt{2}(\bar{f}_1P_R\gamma^{\mu}f_2)(\bar{\chi}_0\gamma_{\mu}\chi^+)
+\right.
\label{dim6op}\\
&&\left.\frac{1}{2}(\bar{f}_1P_R\gamma^{\mu}f_1)\left(
4\bar{\chi}^+P_L\gamma_{\mu}\chi^+-\bar{\chi}_0\gamma^5\gamma_{\mu}\chi_0\right)
+\frac{1}{2}(\bar{f}_2P_R\gamma^{\mu}f_2)\left(
4\bar{\chi}^-P_L\gamma_{\mu}\chi^--\bar{\chi}_0\gamma^5\gamma_{\mu}\chi_0
\right)\right].\nn
\eea
The  annihilation channel into massless fermions ($\chi_0\chi_0\to f\bar f$) is in $p$-wave, as already anticipated,
due to helicity suppression. 
Since we are considering DM  not as a gauge singlet but rather as part of a multiplet carrying a non-zero 
$SU(2)_L$ quantum number, it is possible that a $W$ boson is emitted  from the initial legs of the annihilation process (see ISR diagrams in Fig.~\ref{fig:full}). 
At  ${\cal O}(1/r^2)$ the ISR of gauge boson, which can be computed using the operators in Eq.~(\ref{dim6op}) \citep{paper2}, 
 can lift the helicity suppression and
 its contribution to the $s$-wave cross section is 
 \footnote{
Notice that this is a different scaling with $r$ with respect to the one 
in FSR and VIB, which are ${\cal O}(1/r^4)$ \citep{paper1, paper2}.}
\begin{equation}\label{eq:ISR}
v\sigma_{\rm ISR}(\chi_0\chi_0\to f \bar{f} W)\sim \frac{g^2|y_{\chi}|^4}{M_{\chi}^2}\mathcal{O}\left(\frac{1}{r^2}\right)\mathcal{O}(v^0)\,.
\end{equation} 
where we defined the dimensionless parameter
\begin{equation}
r\equiv\frac{M_{\phi}^2}{M_{\chi}^2}\geq 1\,.
\end{equation} 
Comparing Eq.~(\ref{eq:ISR}) and Eq.~(\ref{eq:WW}) a naive estimation shows that for $y_{\chi}\sim 1$, and in the limit $M_\chi\simeq M_\phi$ the cross sections for DM annihilation  into $f\bar f$
with associated EW radiation can be comparable to, or even bigger than, the one into $W^+W^-$. 
In this regime, the light fermion channel can give a potentially sizeable contribution to the $s$-wave competitive
with the one from di-bosons.
However, in the regime $M_\chi\simeq M_\phi$ the EFT description (\ref{eq:effectiveLagrangian}) is not
reliable anymore and one has to deal with an explicit model for the  interactions between DM and the SM particles. Let us now turn to show the results of a complete calculation within the model presented in section \ref{sec:model}.

\section{Results and discussion}
\label{sec:complete}

\subsection{Cross sections}

\begin{figure}[!htb!]
  \centering
  \includegraphics[width=11 cm]{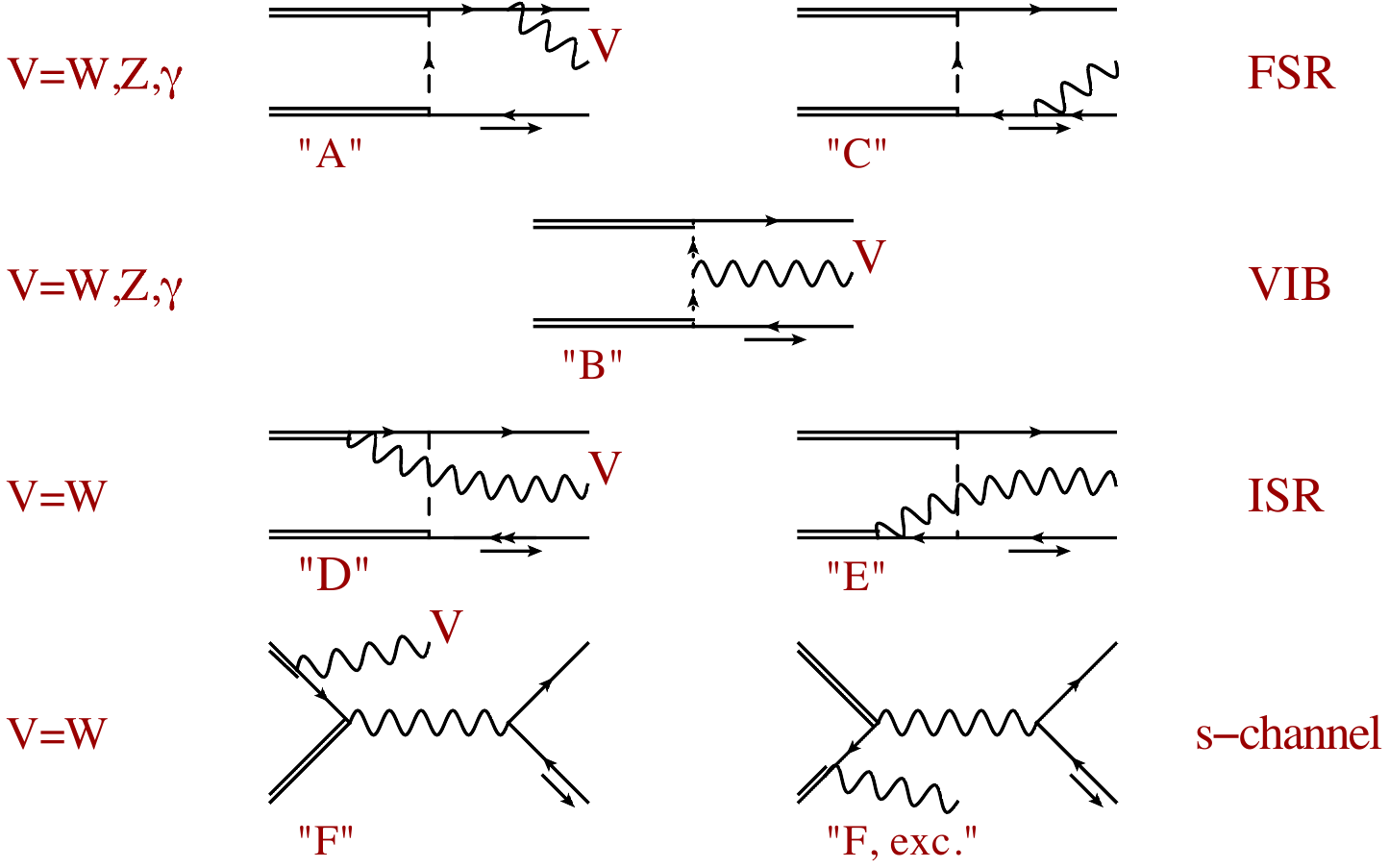}\\
  \caption{\small{
  Feynman diagrams for the annihilation process $\chi_0\chi_0\to f_i\bar{f}_jV$. FSR, VIB, ISR and the s-channel exchange of a 
  charged $W$ boson are present. Notice that for diagrams of type A-E the correspondent ones with crossed initial legs 
  (denoted as ``exc.'' in the expressions of the amplitude collected in Appendix \ref{app:A}) are not shown.}}
  \label{fig:full}
\end{figure}

We consider the 3-body annihilation process
\begin{equation}\label{eq:full3body}
\chi_0(k_1)\chi_0(k_2)\to f_i(p_1)\bar{f}_j(p_2)V(k),
\qquad V=W^{\pm},Z,\gamma;
\end{equation}
the diagrams are shown in Fig.~\ref{fig:full}, where light fermions in the final state can be either leptons or quarks. 
Notice again that -- being DM a neutral particle with zero hypercharge -- the emission of neutral $Z,\gamma$ gauge bosons can occur only through FSR and VIB, leading to the effects already discussed in \citep{bergstrom1,paper1}, while the presence of ISR is a peculiarity of $W$ emission\footnote{In Ref.~\citep{gluonrad}
the emission of a gluon, being the DM colorless, involves again only FSR and VIB.};
therefore for definiteness we present in this Section the explicit result for the cross section considering only the case $\chi_0\chi_0\to e^{+}_L\nu_LW^{-}$, collecting in Appendix \ref{app:A} a compendium of the expressions for the annihilation amplitudes both for leptons and quarks in the final state. 
In the numerical results, we take into account processes where all gauge bosons $\gamma, Z, W^\pm$ are radiated.

Following the discussion in Sec. \ref{sec:eft}, we point out that the various contributions to the amplitude can be schematically organized in the following expression
\begin{align}\label{eq:amplitudesketch}
\mathcal{M}\sim&\frac{g^3}{M_{\chi}}\mathcal{O}(v^0)\,+
\nonumber\\&
\frac{g|y_{\chi}|^2}{M_{\chi}}\left\{
\mathcal{O}(v)\left[\left.\mathcal{O}\left(\frac{1}{r}\right)\right|_{\rm FSR}+
\left.\mathcal{O}\left(\frac{1}{r^2}\right)\right|_{\rm FSR}\right]+\mathcal{O}(v^0)\left[
\left.\mathcal{O}\left(\frac{1}{r}\right)\right|_{\rm ISR}+\left.\mathcal{O}\left(\frac{1}{r^2}\right)\right|_{\rm VIB+FSR}
\right]
\right\}.
\end{align}
In Eq.~(\ref{eq:amplitudesketch}) the first term, corresponding to the $s$-channel gauge boson exchange, and the ISR contribution are present only for 3-body annihilation involving a charged $W$ in the final state.

The cross section for the process $\chi_0\chi_0\to e^+_L\nu_LW^-$  is computed using the  amplitudes reported in Appendix \ref{app:A}.
We are only interested here in the $s$-wave contributions, so we work in the $v\to 0$ limit.
Neglecting all terms suppressed by powers of $m_W/M_\chi$, the 
total cross section reads
\footnote{Although we report only the terms which do not contain powers of $m_W/M_\chi$}, in the numerical computations all terms are included.
\begin{equation}\label{eq:CSapprossimata}
v\sigma(\chi_0\chi_0\to \nu_LW^-e^+_L)=\frac{g^2}{144\pi^3}{|y_{\chi}|^4\over M_{\chi}^2r^2}+
\frac{g^4}{96\pi^3}{|y_{\chi}|^2\over M_{\chi}^2r}+\frac{g^6\left[1+24\ln(2M_{\chi}/m_W)\right]}{4608\pi^3M_{\chi}^2}
+\mathcal{O}\left(\frac{1}{r^3}\right)\,,
\end{equation}
where the first contribution comes from the ISR (diagrams $|{D+E}|^2$), the second involves the interference between ISR and the $s$-channel $W$ exchange [diagrams $({ D+E}){ F}^*+{\rm c.c.}$]  while the third is a pure $s$-channel contribution (diagrams $|{F}|^2$).
For comparison we present here also an analytical approximation for the annihilation cross section related to the process $\chi_0\chi_0\to e^+_Le^-_LZ$, in which only FSR and VIB contribute; we find
\begin{equation}
v\sigma(\chi_0\chi_0\to e_L^+e^-_LZ)= 
\frac{g^2(1-2s_W^2)^{2}}
{3840\pi^{3}c_W^2 }
{|y_\chi|^{4}\over M_{\chi}^{2}r^{4}} + \mathcal{O}\left({1\over r^{5}}\right),
\end{equation}
in the limit $m_{Z}/M_\chi\ll 1$.
A similar formula holds for the cross section of $v\sigma(\chi_0\chi_0\to e_L^+e^-_L\gamma)$,
with an appropriate change of the coefficient, as given in Appendix \ref{app:A}.
We have  though checked that FSR and VIB processes are subleading with respect to W emission in ISR,
in the range of parameters we are considering.

\begin{figure}[t]
   \centering
      \includegraphics[scale=0.7]{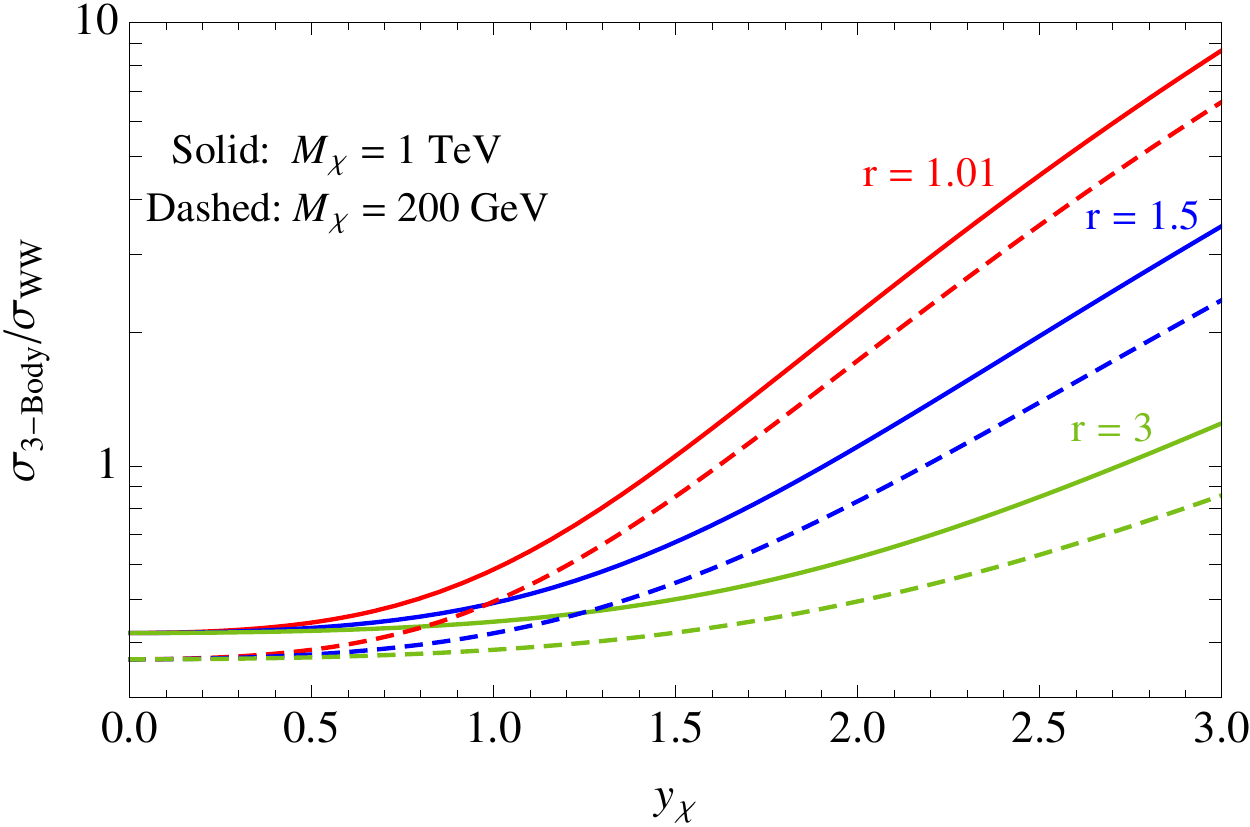}\\
   \includegraphics[scale=0.7]{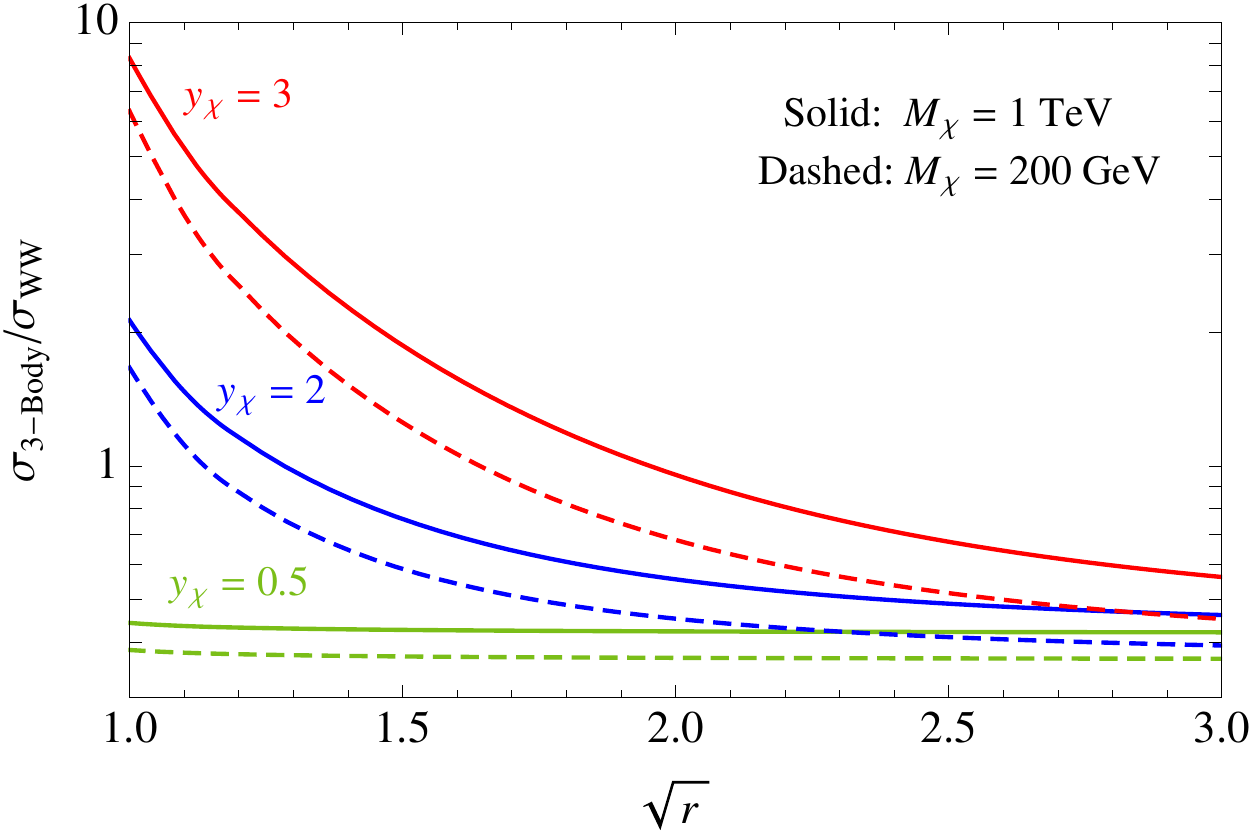}\\
      \includegraphics[scale=0.7]{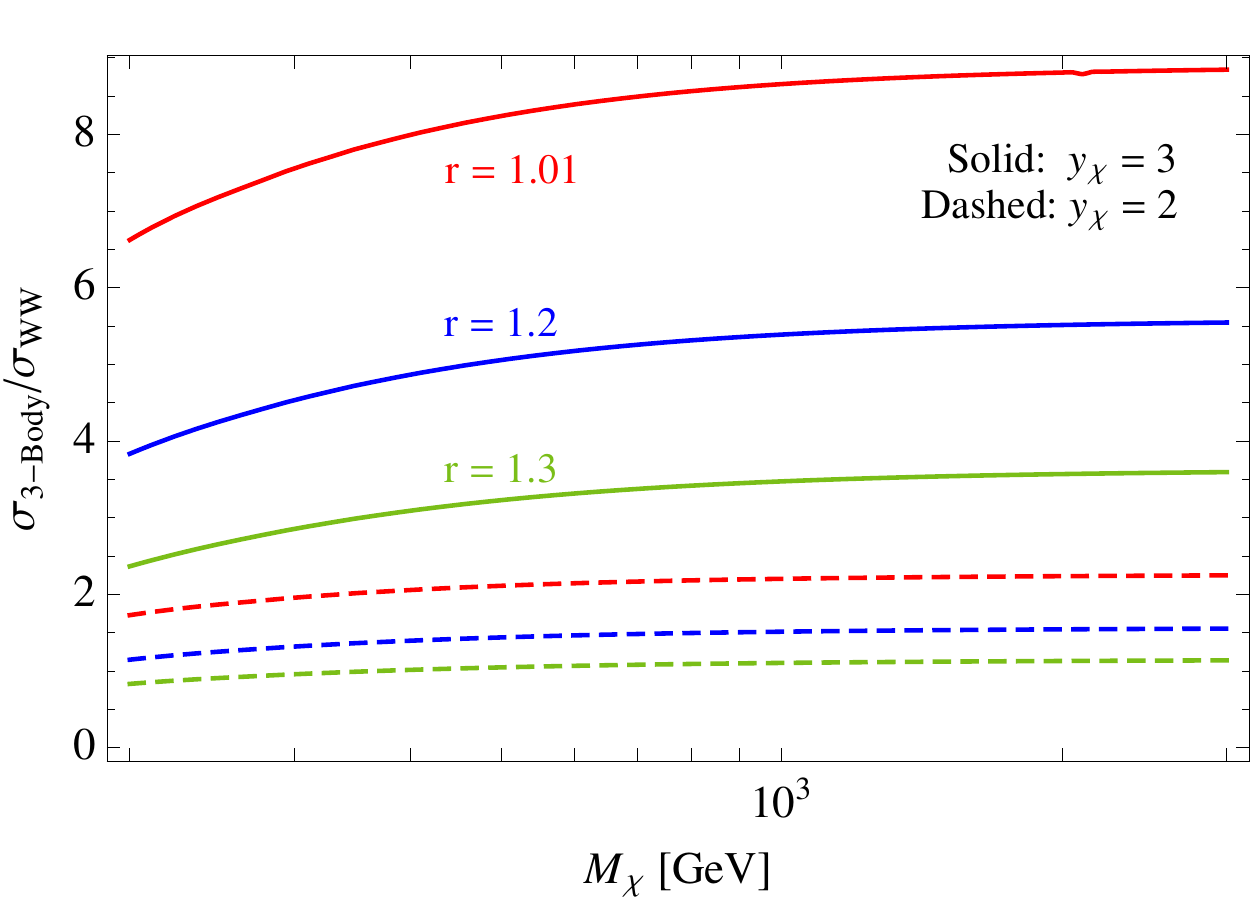}
     \caption{\small{Cross section ratios between the 3-body annihilation processes (two light 
     quarks and $\gamma, Z, W^\pm$ gauge bosons) and the $W^+W^-$ annihilation channel, as a function of $y_\chi$  \emph{(top panel)}, $\sqrt{r}=M_\phi/M_\chi$  \emph{(central panel)}, and $M_\chi$ \emph{(bottom panel)}.
     }}
 \label{fig:CSRatio}
\end{figure}

To show the importance of EW corrections, 
in Fig.~\ref{fig:CSRatio} we show the ratio between the $s$-wave opened by the 3-body annihilations,
summing over the processes in Eq.~(\ref{eq:full3body}), 
and the one from annihilation into $W^+W^-$ for different choices of $M_{\chi}, r$ and $y_\chi$. 
A complete comparison of the cross sections at $\mathcal{O}(g^6)$ would require 
to include also the one-loop electroweak corrections to the $W^+W^-$ channel;
here, we keep for simplicity the tree-level result
(\ref{eq:WW}) as a benchmark for the comparison (see Ref.~\citep{iengo} for 
the full one-loop results).
Increasing $|y_\chi|^2/r$ leads to an increase of the relative importance
of 3-body processes with respect to $W^+ W^-$, while increasing $M_\chi$ makes the EW corrections more effective.
For values of the coupling $y_{\chi}$ of order unity, the 3-body cross section with ISR can be comparable or even dominate over the 2-body annihilation into gauge bosons.
Notice that, in the MSSM, the yukawa coupling is related to the gauge coupling as $y_{\chi}=g/\sqrt{2}\sim 0.5$.
In the limit $y_\chi\to 0$, the 3-body cross section  approaches the value given approximately by the the last term in Eq.~(\ref{eq:CSapprossimata}), which is due to the $s$-channel 
$W$ exchange. 
In the ratio of the cross sections the residual dependence on $M_\chi$ is in the logarithmic 
term and it is larger for larger $M_\chi$. 
As soon as the quantity $|y_\chi|^2/r$ becomes sizeable, the other terms
will  dominate and the $M_\chi$ dependence tends to disappear.
The weak increase of the cross section ratio with $M_\chi$ is apparent from the bottom panel of Fig.~\ref{fig:CSRatio}.

The ratio of the total cross sections is certainly an appropriate estimator of the relative importance of the 3-body
and the di-boson channels. However, the actual observables (the fluxes) are directly related to the
differential cross sections per unit of energy. As we are going to show in the next subsection,
even in situations where 
$\sigma(\chi_0\chi_0\to \nu_LW^-e^+_L)<\sigma(\chi_0\chi_0\to W^+W^-)$,
the former process can contribute much more significantly than the latter one to the energy spectra of 
final particles, in given energy bands.

\subsection{Energy spectra of final stable particles}

The analytical calculations of the diagrams in Fig.~\ref{fig:full}
can be used to derive the energy spectra of stable SM particles originated by DM annihilation events, at the interaction point, i.e. before the astrophysical propagation to Earth. 
This part necessarily involves numerical  techniques to simulate a large number of annihilation events, with the
inclusion of EW radiation, and  then evolve them according to SM evolution, including QCD showering, hadronization and decays.
The simulations were carried out with our own Monte Carlo code (interfaced to {\sc Pythia 8} \citep{pythia}), as already explained
in Refs.~\citep{paper1,paper2}.

Working  in the approximation of massless external fermions and zero relative velocity we will focus on the case in which the DM triplet is coupled either to the lepton or 
the quark sector, for which the primary annihilation channels for $\chi_0\chi_0 \to {\rm I}$ (including EW bremsstrahlung) are
\bea
\label{eq:leptons}
{\rm I}_{\rm leptons}&=&\left\{
W^{+}W^{-},~e^{+}_Le^{-}_L\gamma,~e^{+}_Le^{-}_LZ,~\nu_L\bar{\nu}_{e\,L}Z,
~e^{+}_L\nu_{e\,L}W^{-},~e^{-}_L\bar{\nu}_{e\,L}W^{+}
\right\}\,, \\
\label{eq:quarks}
{\rm I}_{\rm quarks}&=&\left\{
W^{+}W^{-},~u_L\bar{u}_L\gamma,~d_L\bar{d}_L\gamma,~u_L\bar{u}_LZ,
~d_L\bar{d}_{L}Z,
~u_L\bar{d}_{L}W^{-},~d_L\bar{u}_{L}W^{+}
\right\}\,,
\eea
respectively.
Once again notice that the contribution of ISR affects only the channels with the emission of a $W$. For those involving the $\gamma, Z$ emission,
through FSR and VIB, we use the results of Ref.~\citep{paper1}.  In \citep{paper1},  in order to have also an estimation of the $p$-wave contribution,  the annihilation cross sections were evaluated considering $v\neq 0$ and  it was performed an expansion up to the order $\mathcal{O}(1/r^4)$. On the contrary, we work here in the limit $v\to 0$,  so we can use  
the exact analytical expressions.

\begin{figure}[t]
\begin{center}
   \includegraphics[scale=0.5]{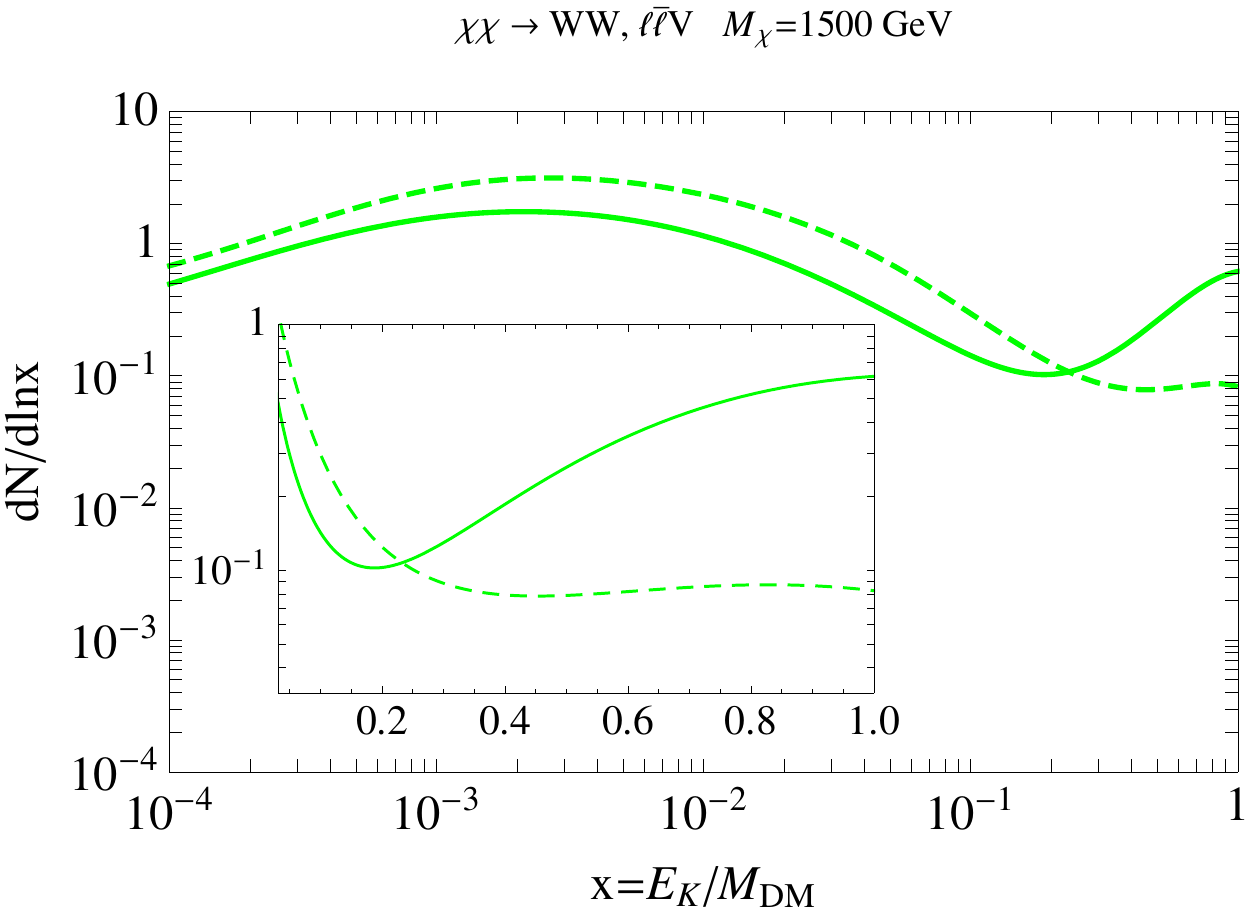}
   \hspace{1cm}
      \includegraphics[scale=0.5]{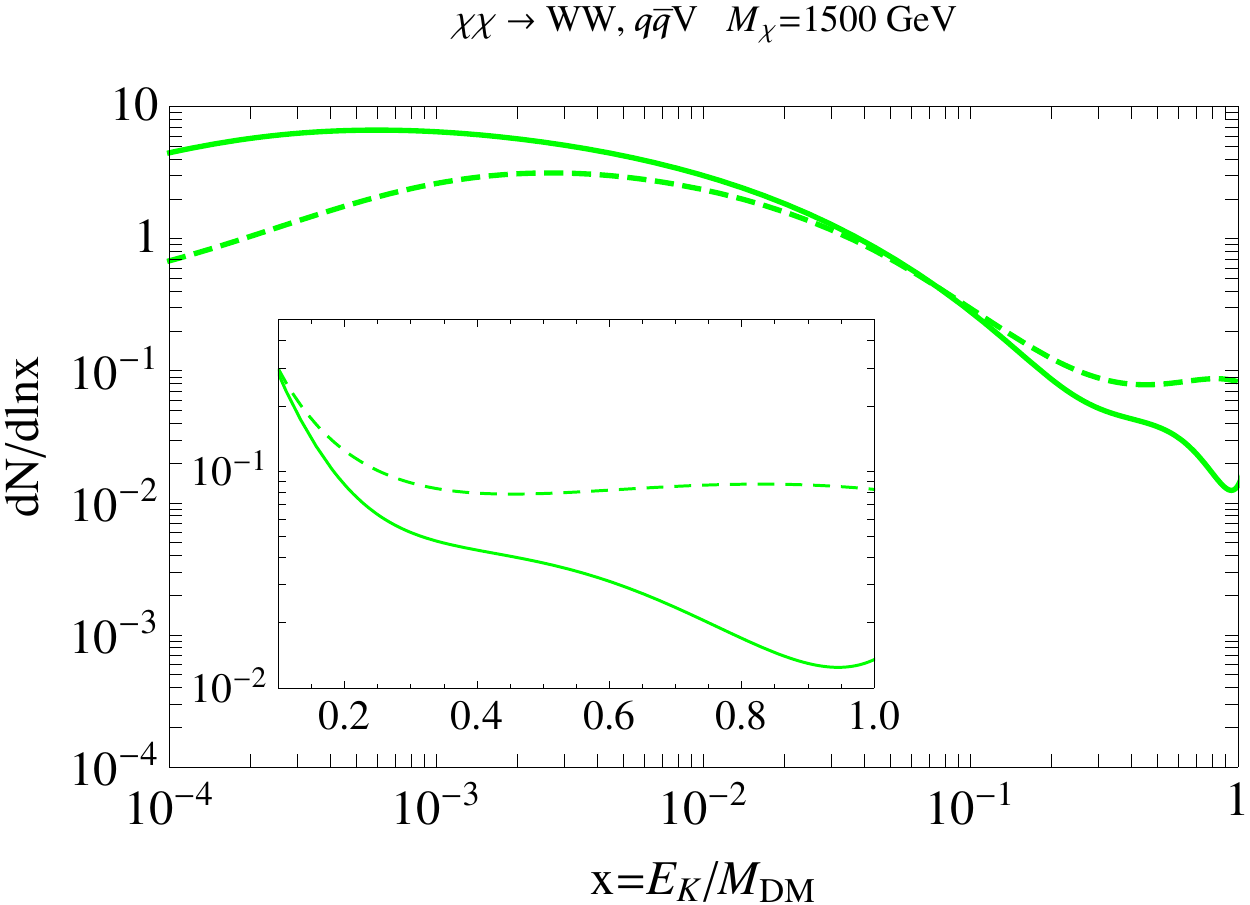}\\
   \includegraphics[scale=0.5]{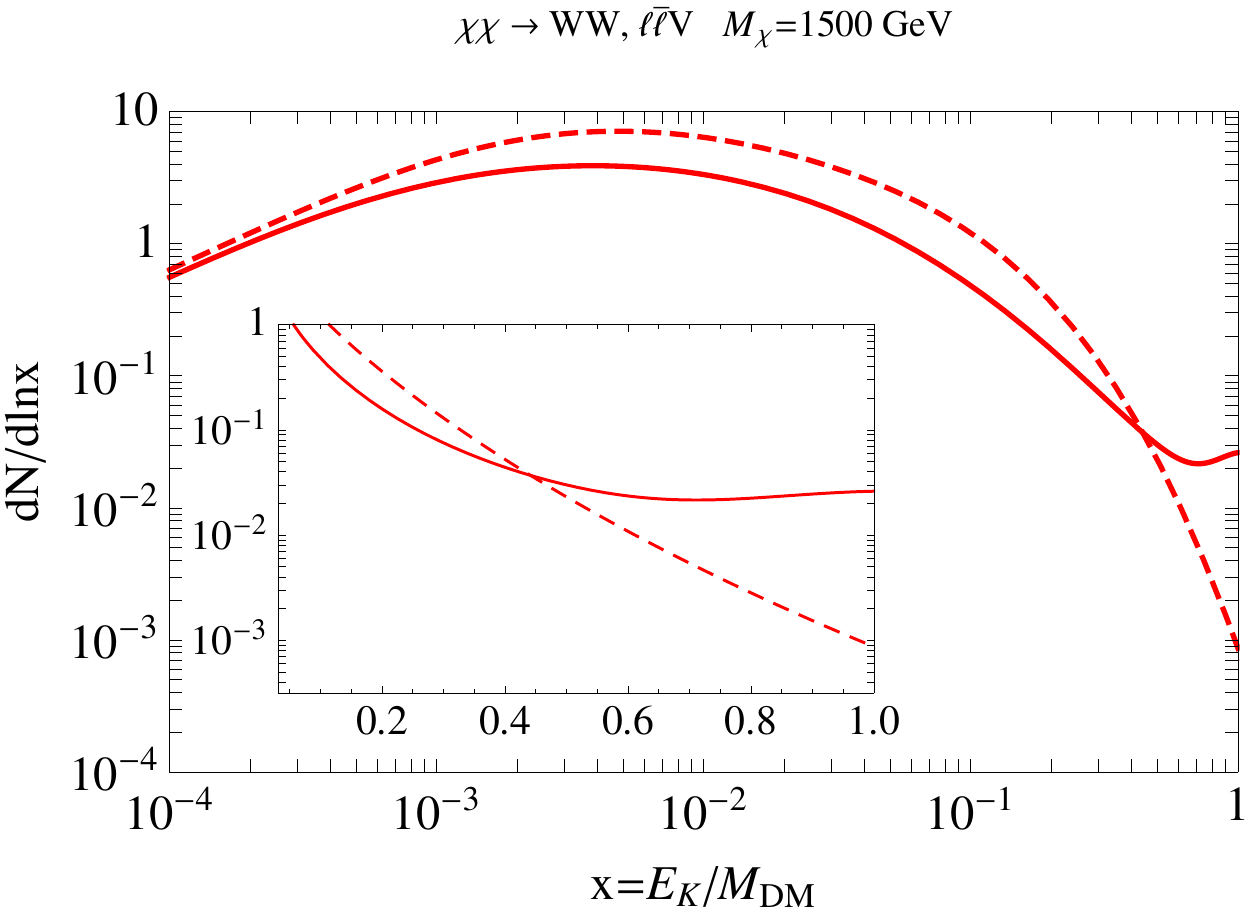}
      \hspace{1cm}
   \includegraphics[scale=0.5]{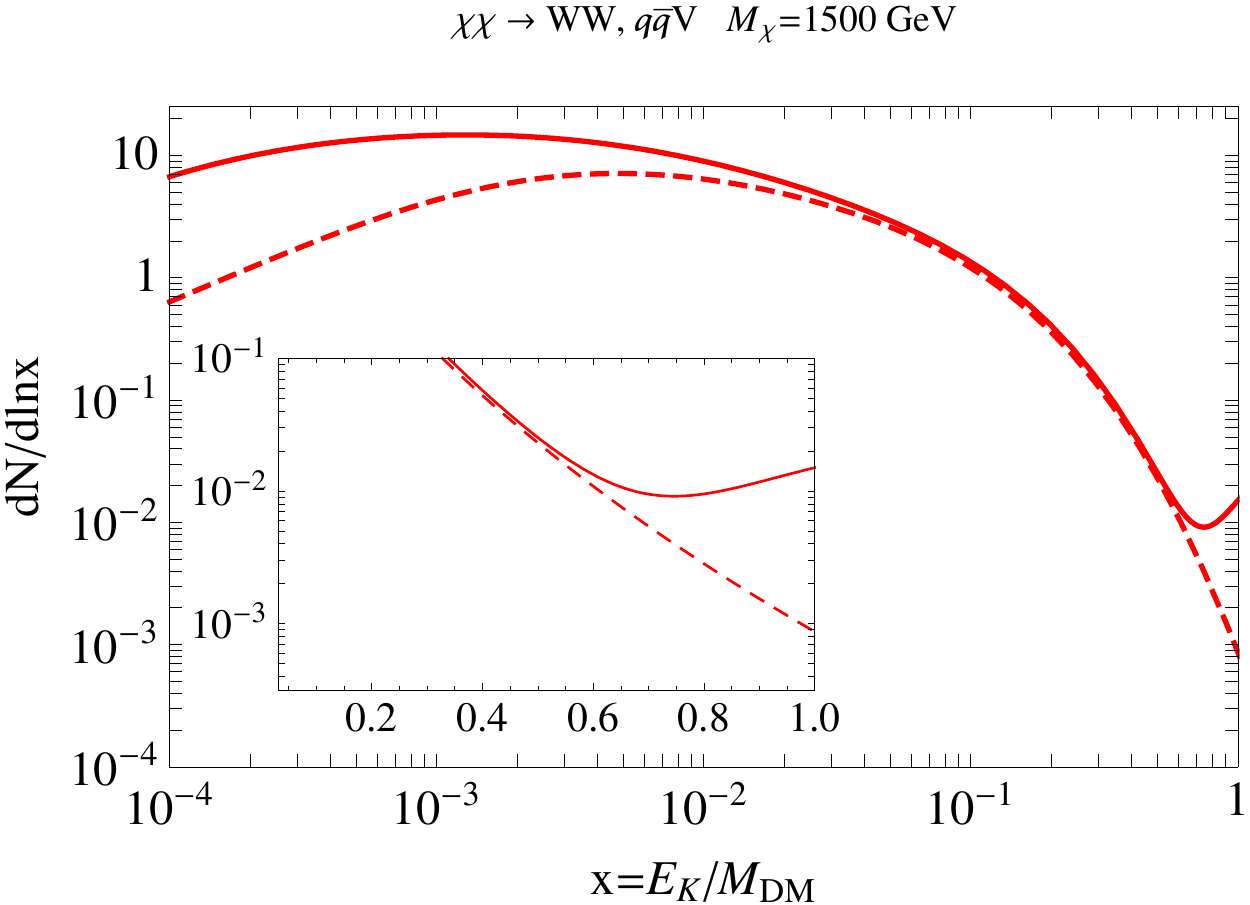}\\
   \includegraphics[scale=0.5]{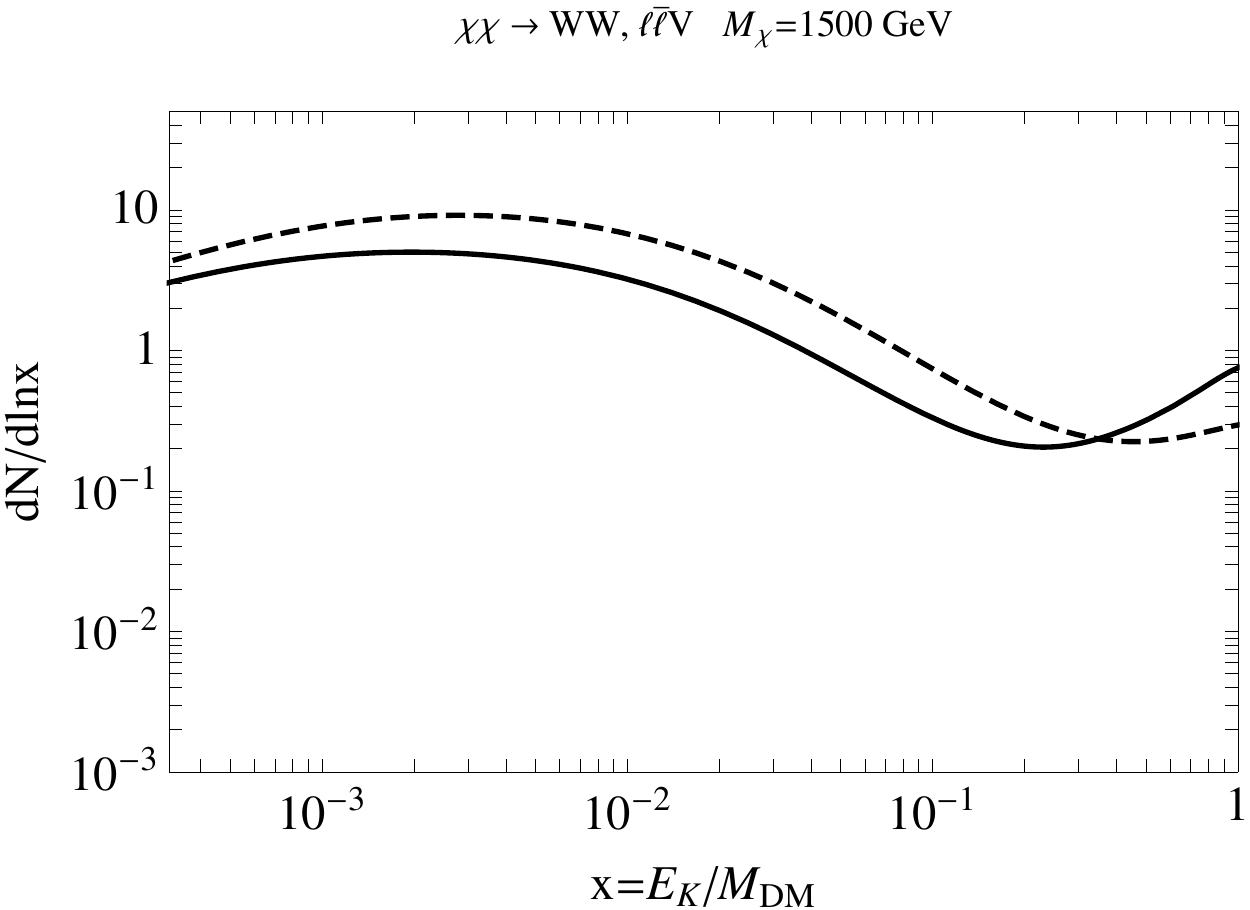}
      \hspace{1cm}
   \includegraphics[scale=0.5]{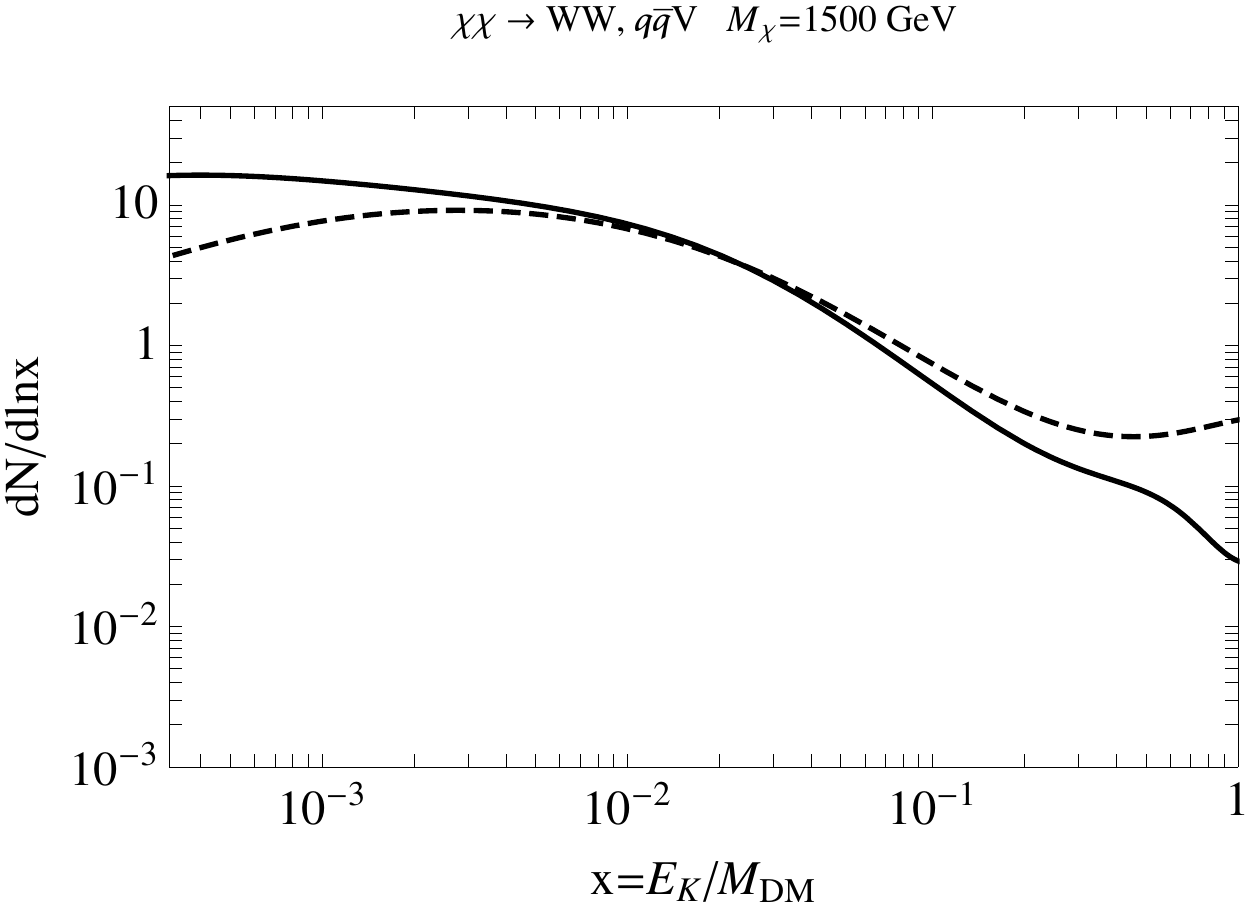}\\
     \includegraphics[scale=0.5]{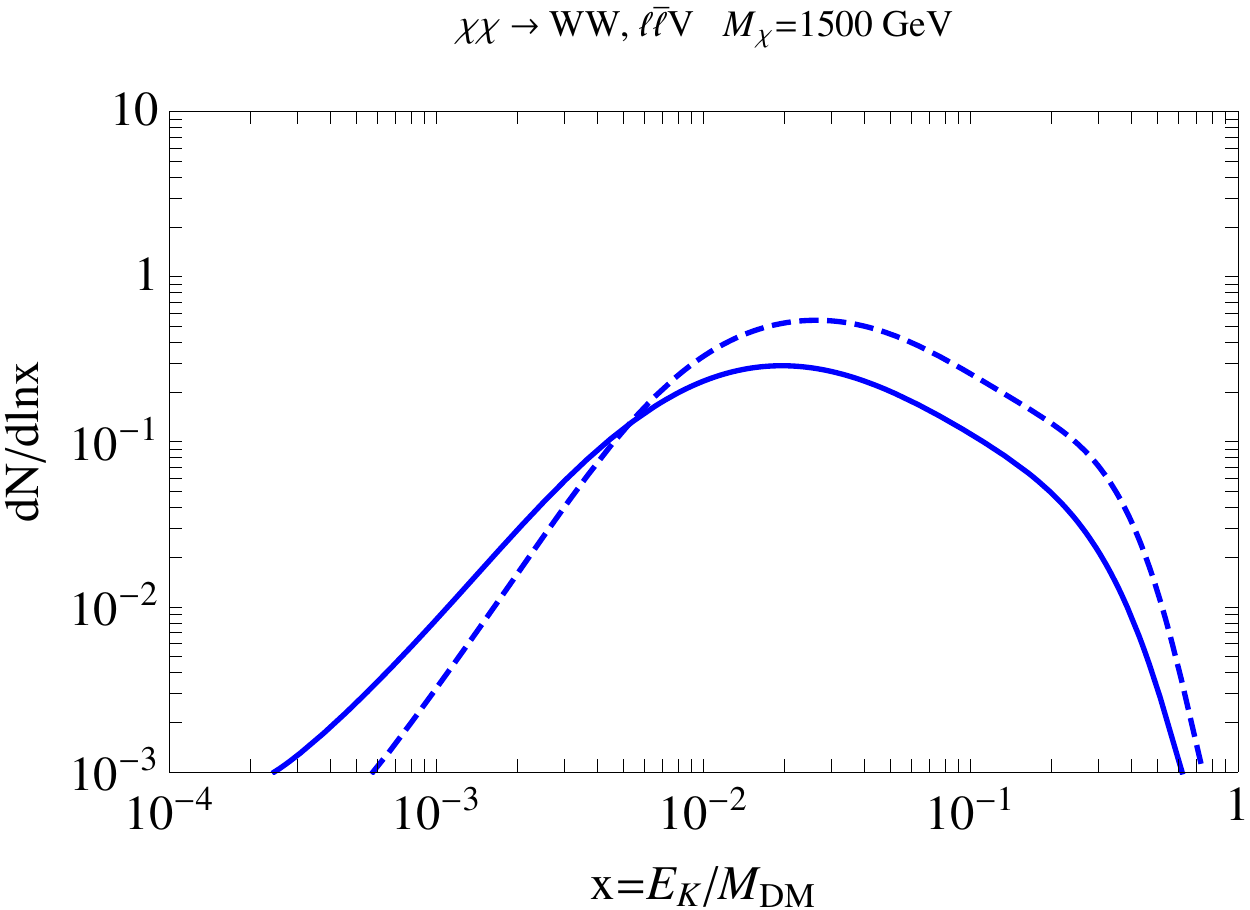} 
      \hspace{1cm}
      \includegraphics[scale=0.5]{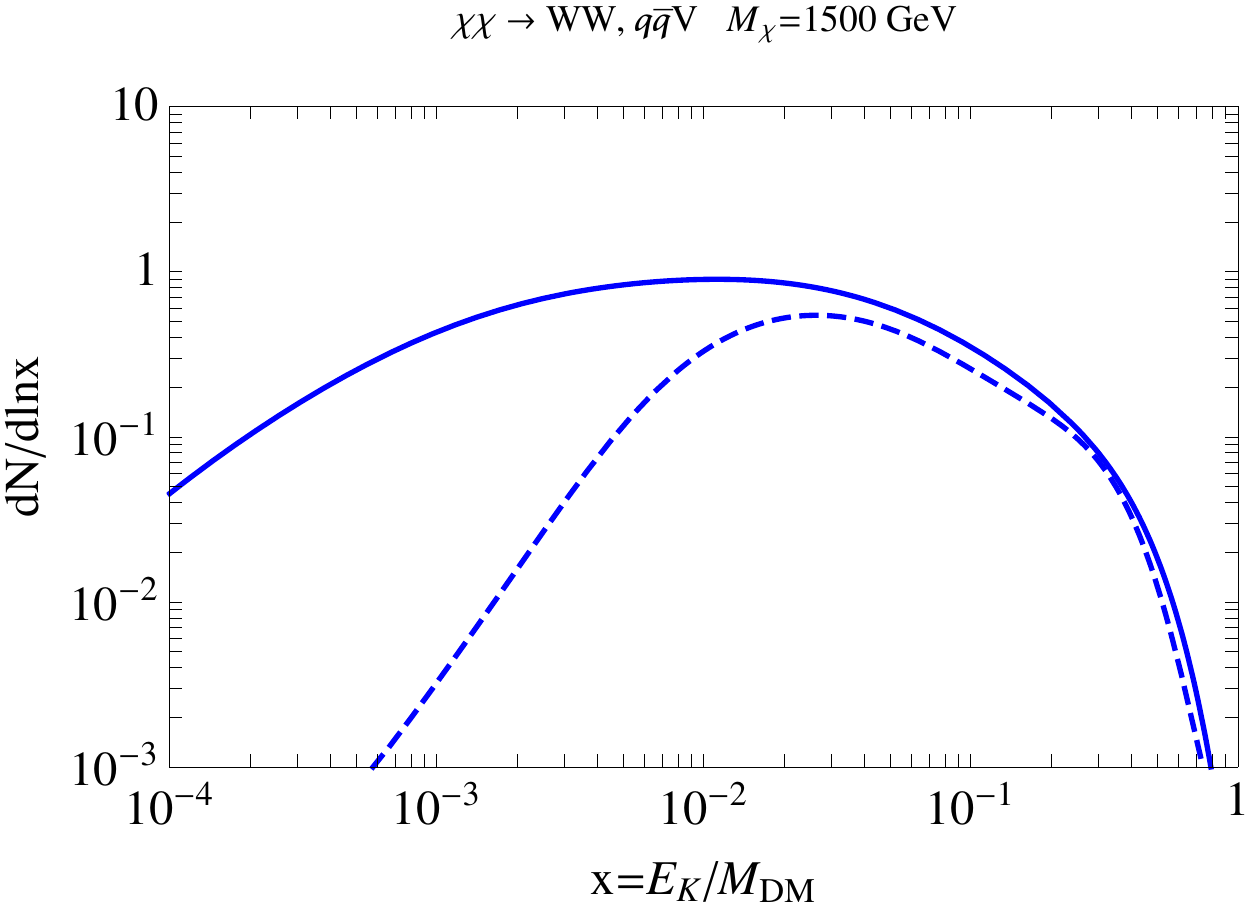}   
 \caption{\small{
     Energy spectra of final $e^+$ (green), $\gamma$ (red), $\nu_e+\nu_\mu+\nu_\tau$ (black), $\bar p$ (blue)
     for the cases of a lepton final state  (left column) or a quark final state  (right column).
     Parameters are set as $M_\chi=1.5 \tev, r=1.1, y_\chi=2$. For comparison,
     the spectra from the $W^+W^-$ channel are also shown (dashed lines), 
     corresponding to $y_\chi=0$.
     }}
 \label{fig:spectra1}
 \end{center} 
\end{figure}

From the numerical results, we extract the energy distributions
of each stable species $f$
\be
\label{eq:norm}
\frac{d{N}_f}{d \ln x}\equiv {1\over\sigma(\chi_0\chi_0\to \rm I)}{d\sigma(\chi_0\chi_0\to f+X)\over d\ln x}
 \,,\qquad f=\{ e^+, e^-, \gamma, \nu,\bar\nu, p,\bar p\}\,,
\ee
where $x\equiv E_{\rm kinetic}^{(f)}/M_{\chi}$, $E_{\rm kinetic}^{(f)}$ is the kinetic energy of the particle $f$ (distinguishing between total and kinetic energies is obviously 
relevant only for (anti)protons), and the  $X$ reminds us of the inclusivity in the final state with respect to the particle $f$.
The  normalization in (\ref{eq:norm}) is $\sigma(\chi_0\chi_0\to\rm I)$, where $\rm I=\rm{I}_{\rm leptons}, \rm{I}_{\rm quarks}$, and stands for the sum of all open annihilation channels.
In Fig.~\ref{fig:spectra1} we plot the  resulting ${d{N}_f}/{d x}$  for $e^+, \gamma, \nu=(\nu_e+\nu_\mu+\nu_\tau)/3, 
\bar p$ for specific, but representative, choices of parameters, as originating either from fermion channels with ISR (solid lines) or from  $W^+ W^-$ (dashed lines). 
We set the model parameters as $y_\chi=2$, $M_\chi=1.5$ TeV and $r=1.1$;
the corresponding  ratios of the total cross sections are $\sigma_{\rm 3-body}/\sigma_{WW}\simeq 0.6$,
for the lepton channel, and $\sigma_{\rm 3-body}/\sigma_{WW}\simeq 1.8$ for the quark channel.

We consider two different channels, the lepton and quark channels defined in (\ref{eq:leptons})-(\ref{eq:quarks}).
The spectra due to ISR result to be competitive with those obtained from $W^+W^-$.
Interesting features  can be extracted from this comparison.
For the lepton  channel, the very hard positrons  can be an order of magnitude more abundant for ISR than for $W^+W^-$,
due to the contribution of the primary positrons. 
On the other hand, in the quark channel all final species originate from radiative and/or hadronization processes,
except for the primary photons from $f \bar f \gamma$ (FSR and VIB diagrams); therefore, the resulting spectra
are largely dominated by the soft regions $x\lesssim 10^{-2}$. 
This fact is even more apparent  for the antiprotons,  which are copiously generated by the soft $W$ emitted in ISR
and by the hadronization of the primary quarks,
and they can easily overcome the antiprotons produced in the di-boson channel by two or more orders of magnitude.
For both cases of quark and lepton channels, the primary photons from $f \bar f \gamma$  enhance the spectrum of hard gamma-rays by one or two
orders of magnitude with respect to that from annihilations into $W^+W^-$.
This  peak of the gamma-ray spectrum  in the hard region can have very important 
implications for actual searches (see e.g. Ref.~\citep{Bringmann:2011ye} for an analysis of gamma-ray  features near the spectrum endpoint).

As already anticipated in the previous subsection, 
 although the ratio of the total cross sections $\sigma_{\rm 3-body}/\sigma_{WW}$ is around (or even smaller than) 1, as in the case of the MSSM wino, the resulting energy spectra can differ by 
orders of magnitude, in some energy bands. 

In order to derive from these results the real phenomenological observables one would need to integrate the energy spectra over the diffused
source (the DM distribution in the halo) and then propagate them from the annihilation region to Earth, according to the standard equations of cosmic-ray evolution.
We expect the pronounced features in the hard regions of the gamma-ray and positron spectra to be preserved after distribution and propagation uncertainties are taken into account.

\section{Conclusions and outlook}
\label{sec:conclusions}

In this paper we have addressed the question of how important is the effect of  EW bremsstrahlung
for the annihilation of DM particles in the galactic halo, assuming that the DM is the neutral Majorana
component of an $SU(2)_L$-triplet, like a wino in the MSSM. 
Our analysis complements that of Ref.~\citep{paper2}, which was carried out in the context of an EFT,
in that it applies to a regime where the EFT is not reliable, namely when the DM mass  (setting the 
scale of the annihilation process) is very close to the cutoff scale of the effective theory. 
Actually, this is precisely the regime where the effects of EW bremsstrahlung are more pronounced.
To study this regime, we resorted to make calculations in the context of a simple model corresponding 
to the MSSM interactions of a wino with SM fermions and their scalar partners.

The DM annihilates predominantly into $W^+W^-$, if kinematically allowed. However, the initial
state of the annihilation process into light fermions can radiate an EW boson; this process lifts the 
helicity suppression and contributes to the $s$-wave cross section.
We found that, in a large portion of the parameter space the total cross section into light fermions (with
the inclusion of ISR) $\sigma_{\rm 3-body}$ is comparable to that into di-bosons $\sigma_{WW}$, and there are even situations where $\sigma_{\rm 3-body}>\sigma_{WW}$.
This happens when the parameter $|y_\chi|^2/r$ becomes sizeable, see e.g. Fig.~\ref{fig:CSRatio}.
However, the energy spectra are sensitive to the differential cross sections.
Thus, even in situations where $\sigma_{\rm 3-body}<\sigma_{WW}$ (e.g. in the case of the MSSM wino) the  spectra of final particles can differ by orders of magnitude, in some energy bands.
For example,  hard positrons, hard gamma rays and soft antiprotons coming from processes with fermion final states
and EW bremsstrahlung can be one or two orders of magnitude more abundant than those originating
from the di-boson channel.
We conclude that EW bremsstrahlung must be taken into account in order to make reliable predictions 
for indirect DM searches.

Our results can be applied to place limits on the parameters of the model, and hence on the MSSM
with wino DM, by e.g.~imposing the non-observation of excesses in existing data sets of gamma-ray
or antiproton fluxes.
For a pure MSSM wino,  we expect that the correct relic abundance is obtained for a value of the DM mass 
in the TeV-range (the actual value is 2.7 TeV in the case of heavy sfermions and including
electroweak Sommerfeld effects
\citep{winomass}), while the spin-independent cross section per nucleon is about $10^{-45} \textrm{ cm}^2$
\citep{mdm}. 
This situation is beyond the scope of LHC-7 searches and of
ongoing direct detection experiments. Therefore, indirect detection represents the most promising
tool to probe this kind of DM candidates, in the near future.
In this perspective, a fruitful technique is to study the correlations among the fluxes of the different
species, originating from DM annihilations with the inclusion of EW bremsstrahlung;
this way, it would be possible to compare observations from different indirect detection 
experiments and to test the hypothesis that the putative signals have actually a DM origin
\citep{futuro}.

\acknowledgments
We thank Marco Cirelli for useful discussions and comments on the manuscript.
The work of A.U. is supported by the program \'Emergence-UPMC-2012.

\appendix

\section{Matrix elements for the 3-body annihilation}
\label{app:A}

We collect here the annihilation amplitudes for the Feynman diagrams in Fig.~\ref{fig:full} related to the processes
\begin{equation}
\chi_0(k_1)\chi_0(k_2)\to f_i(p_1)\bar{f}_j(p_2)V(k).
\end{equation}
Explicit expressions are presented for the case $\chi_0\chi_0\to e^+_LW^-\nu_L$ (equal to its CP conjugated); 
because of the massless limit for final light fermions, in fact, all the other ones - both for leptons and quarks - can be obtained with a straightforward manipulation of the couplings (see Tab. \ref{tab:coeff}).\\
Each amplitude consists in two contributions related by the exchange of the initial Majorana particles ($k_1\leftrightarrow k_2$, denoted as ``exc.'') with a minus sign due to Fermi 
statistics; defining $\mathcal{M}_{k}\equiv 
\mathcal{M}_{k}-\mathcal{M}_{k}^{\rm exc.}$, $k=A\dots F$,  
introducing the short-hand notation
\begin{equation}\label{eq:sh}
D_{ij}\equiv \frac{1}{M_{\chi}^2-2p_i\cdot k_j-M_{\phi}^2},
\end{equation}
and neglecting the mass splitting between the neutral and the charged component in the DM multiplet, we find for the amplitude - after a Fierz transformation - the following contributions:

\begin{eqnarray}
\mathcal{M}_{A}&=&\frac{-g|y_{\chi}|^2\left[
\bar{u}_{\nu}(p_1)\slashed{\epsilon}^*(\slashed{k}+\slashed{p}_1)P_R\gamma^{\rho}v_{e}(p_2)
\right]}{
2\sqrt{2}(m_{W}^2+2k\cdot p_1)}
\bar{v}_{\chi}(k_2)\left(D_{22}P_L\gamma_{\rho}-D_{21}\gamma_{\rho}P_L\right)u_{\chi}(k_1),\label{eq:A}\\
\mathcal{M}_{B}&=&\frac{g|y_{\chi}|^2\left[\bar{u}_{\nu}(p_1)P_R\gamma^{\rho}v_{e}(p_2)\right]}
{2\sqrt{2}}\bar{v}_{\chi}(k_2)\left[D_{11}D_{22}
\epsilon^*\cdot(k_2-k_1+p_1-p_2)P_L\gamma_{\rho}\right.
\nonumber\\
&&-
\left.D_{12}D_{21}
\epsilon^*\cdot(k_1-k_2+p_1-p_2)\gamma_{\rho}P_L\right]u_{\chi}(k_1),\label{eq:B}\\
\mathcal{M}_{C}&=&\frac{g|y_{\chi}|^2\left[
\bar{u}_{\nu}(p_1)P_R\gamma^{\rho}(\slashed{k}+\slashed{p}_2)\slashed{\epsilon}^*v_{e}(p_2)
\right]}{
2\sqrt{2}(m_{W}^2+2k\cdot p_2)}
\bar{v}_{\chi}(k_2)\left(D_{11}P_L\gamma_{\rho}-D_{12}\gamma_{\rho}P_L\right)u_{\chi}(k_1),\label{eq:C}\\
\mathcal{M}_{D}&=&\frac{-g|y_{\chi}|^2\epsilon_{\mu}^{*}(k)\left[\bar{u}_{\nu}(p_1)P_R\gamma^{\rho}v_{e}(p_2)\right]}
{\sqrt{2}}\times\nonumber\\&&
\bar{v}_{\chi}(k_2)\left[
\frac{D_{22}P_L\gamma_{\rho}(\slashed{k}_1-\slashed{k}+M_{\chi})\slashed{\epsilon}^*}{(m_W^2-2k_1\cdot k)}
-
\frac{D_{21}\slashed{\epsilon}^*(\slashed{k}_2-\slashed{k}-M_{\chi})\gamma_{\rho}P_L}{(m_W^2-2k_2\cdot k)}
\right]u_{\chi}(k_1),\label{eq:D}\\
\mathcal{M}_{E}&=&\frac{-g|y_{\chi}|^2\left[\bar{u}_{\nu}(p_1)P_R\gamma^{\rho}v_{e}(p_2)\right]}
{\sqrt{2}}\times\nonumber\\&&
\bar{v}_{\chi}(k_2)\left[
\frac{D_{11}\slashed{\epsilon}^*(\slashed{k}-\slashed{k}_2+M_{\chi})P_L\gamma_{\rho}}{(m_W^2-2k_2\cdot k)}
-
\frac{D_{12}\gamma_{\rho}P_L(\slashed{k}-\slashed{k}_1-M_{\chi})\slashed{\epsilon}^*}{(m_W^2-2k_1\cdot k)}
\right]u_{\chi}(k_1),\label{eq:E}\\
\mathcal{M}_{F}&=&\frac{g^3\left[\bar{u}_{\nu}(p_1)P_R\gamma^{\rho}v_{e}(p_2)\right]}
{\sqrt{2}(2p_1\cdot p_2-m_W^2+i\Gamma_W m_W)}\times\nonumber\\&&
\bar{v}_{\chi}(k_2)\left[
\frac{\gamma_{\rho}(\slashed{k}_1-\slashed{k}+M_{\chi})\slashed{\epsilon}^*}{(m_W^2-2k_1\cdot k)}
-
\frac{\slashed{\epsilon}^*(\slashed{k}_2-\slashed{k}-M_{\chi})\gamma_{\rho}}{(m_W^2-2k_2\cdot k)}
\right]u_{\chi}(k_1).\label{eq:F}
\end{eqnarray}

\begin{table}[t]
\begin{center}
\begin{tabular}{| c | c | c | c |}
\hline\hline
$\chi_0\chi_0\to {\rm I} $& ${\rm I}=\left\{\begin{array}{c}
 e^+_Le^-_LZ\vspace{1 mm} \\
 d_L\bar{d}_LZ
 \end{array}\right.$ & ${\rm I}=\left\{\begin{array}{c}
 \nu_L\bar{\nu}_LZ \vspace{1 mm}\\
 u_L\bar{u}_LZ
 \end{array}\right.$ &  ${\rm I}=\left\{\begin{array}{c}
 e^+_Le^-_L\gamma \vspace{1 mm}\\
 u_L\bar{u}_L\gamma\vspace{1 mm}\\
 d_L\bar{d}_L\gamma
 \end{array}\right.$ \\ 
  \hline\hline
 $\mathcal{M}_{ A}\to \#\,\mathcal{M}_{ A}$ & $-(1+2Q_{e,d}s_W^2)/\sqrt{2}c_W$ & $+(1+2Q_{e,d}s_W^2)/\sqrt{2}c_W$ & $ -(1+2Q_{e,d}s_W^2)/\sqrt{2}c_W$ \\
 \hline
 $\mathcal{M}_{ B}\to \#\,\mathcal{M}_{B}$ & $+(1+2Q_{\nu,u}s_W^2)/\sqrt{2}c_W$ & 
 $ -(1+2Q_{\nu,u}s_W^2)/\sqrt{2}c_W$ & $+(1+2Q_{\nu,u}s_W^2)/\sqrt{2}c_W$ \\
 \hline
 $\mathcal{M}_{ C}\to \#\,\mathcal{M}_{ C}$ & $+Q_{e,u,d}\sqrt{2}s_W$ & $ -Q_{e,u,d}\sqrt{2}s_W$ & $+Q_{e,u,d}\sqrt{2}s_W$ \\
\hline
 $\mathcal{M}_{D,E,F}\to \#\,\mathcal{M}_{ D,E,F}$ & 0 &  0 & 0 \\
 \hline
  \end{tabular}
  \end{center}
  \caption{\small{Relations between the scattering amplitudes $\mathcal{M}_{A,\cdots,F}$ for the process $\chi_0\chi_0\to e_L^+ W^- \nu_L$ and  
  those for the other processes with $Z,\gamma$ radiation.
  The values for the amplitude $\mathcal{M}_{A\dots F}$ of each diagram are reported in Eqs. (\ref{eq:A})-(\ref{eq:F}). The charges are
  $Q_e=-1$, $Q_{\nu}=0$, $Q_{u}=2/3$ and $Q_d=-1/3$.
  }}\label{tab:coeff}
\end{table}

\noindent
Notice that we check the consistence of these expression using the Ward Identities for EW SM gauge bosons, namely $k_{\mu}\mathcal{M}^{\mu}_L\sim 0$ for $m_f\sim 0$, where $\mathcal{M}_{L}^{\mu}$ is the amplitude computed for the longitudinal mode of the $W$.

Concerning the amplitudes related to the others 3-body final states, we first notice that for the quark channels  
$\chi_0\chi_0\to u_L\bar {d_L}W^-$ and $\chi_0\chi_0\to d_L\bar{u}_LW^+$ the amplitude for the various diagrams is exactly the same, as dictated by isospin invariance; for the other processes, defining the short-hand notation $s_W\equiv\sin\theta_W,\,c_W\equiv\cos\theta_W$, we  find the relations sketched in Tab.~\ref{tab:coeff}. Notice, moreover, that for final states involving light quarks the cross sections get an extra color factor $N_{C}=3$.


\bibliographystyle{JHEP}

\begin{thebibliography}{10}

\bibitem{PPPC} 
  M.~Cirelli, G.~Corcella, A.~Hektor, G.~Hutsi, M.~Kadastik, P.~Panci, M.~Raidal and F.~Sala {\it et al.},
  JCAP {\bf 1103}, 051 (2011)
  	  \href{http://arXiv.org/abs/1012.4515}{[arXiv:1012.4515]}.
  
\bibitem{paper0}
  P.~Ciafaloni, D.~Comelli, A.~Riotto, F.~Sala, A.~Strumia and A.~Urbano,
  JCAP {\bf 1103}, 019 (2011)
	  \href{http://arXiv.org/abs/1009.0224}{[arXiv:1009.0224]}.

\bibitem{bell1} 
N.~F.~Bell, J.~B.~Dent, T.~D.~Jacques and T.~J.~Weiler,
  Phys.\ Rev.\  D {\bf 83}, 013001 (2011)
                 \href{http://arXiv.org/abs/1009.2584}{[arXiv:1009.2584]};
 N.~F.~Bell, J.~B.~Dent, T.~D.~Jacques and T.~J.~Weiler,
  Phys.\ Rev.\ D {\bf 84}, 103517 (2011)
                   \href{http://arXiv.org/abs/1101.3357}{[arXiv:1101.3357]}.

\bibitem{paper1}
  P.~Ciafaloni, M.~Cirelli, D.~Comelli, A.~De Simone, A.~Riotto and A.~Urbano,
  JCAP {\bf 1106}, 018 (2011) 
  \href{http://arXiv.org/abs/1104.2996}{[arXiv:1104.2996]}.

\bibitem{bell2} 
  N.~F.~Bell, J.~B.~Dent, A.~J.~Galea, T.~D.~Jacques, L.~M.~Krauss and T.~J.~Weiler,
  Phys.\ Lett.\ B {\bf 706}, 6 (2011)
  \href{http://arXiv.org/abs/1104.3823}{[arXiv:1104.3823]}.
  
\bibitem{cheung} 
  K.~Cheung, P.~-Y.~Tseng and T.~-C.~Yuan,
  JCAP {\bf 1106}, 023 (2011)
                   \href{http://arXiv.org/abs/1104.5329}{ [arXiv:1104.5329]}.

\bibitem{ibarra1}
  M.~Garny, A.~Ibarra and S.~Vogl,
  JCAP {\bf 1107}, 028 (2011)
                   \href{http://arXiv.org/abs/1105.5367}{[arXiv:1105.5367]}.
  
\bibitem{paper2}
  P.~Ciafaloni, M.~Cirelli, D.~Comelli, A.~De Simone, A.~Riotto, A.~Urbano,
  JCAP {\bf 1110}, 034 (2011)
   \href{http://arXiv.org/abs/1107.4453}{[arXiv:1107.4453]}.

\bibitem{iengo} 
  A.~Hryczuk and R.~Iengo,
                   \href{http://arXiv.org/abs/1111.2916}{arXiv:1111.2916}.
  
\bibitem{barger} 
  V.~Barger, W.~-Y.~Keung and D.~Marfatia,
                   \href{http://arXiv.org/abs/1111.4523}{arXiv:1111.4523}.

\bibitem{ibarra2} 
  M.~Garny, A.~Ibarra and S.~Vogl,
                   \href{http://arXiv.org/abs/1112.5155}{arXiv:1112.5155}.  

\bibitem{proc}
  A.~De Simone,
                   \href{http://arXiv.org/abs/1201.1443}{arXiv:1201.1443}.
  
\bibitem{bergstrom1}
L.~Bergstrom,
  Phys.\ Lett.\  B {\bf 225}, 372  (1989).
\bibitem{bergstrom2}
T.~Bringmann, L.~Bergstrom and J.~Edsjo,
  JHEP {\bf 0801}, 049 (2008) 
	  \href{http://arXiv.org/abs/arXiv:0710.3169} {[arXiv:0710.3169]};
L.~Bergstrom, T.~Bringmann and J.~Edsjo,
  Phys.\ Rev.\  D {\bf 78},  103520 (2008)
	  \href{http://arXiv.org/abs/0808.3725}{[arXiv:0808.3725]}.

\bibitem{list1}
V.~Berezinsky, M.~Kachelriess and S.~Ostapchenko,
Phys.\ Rev.\ Lett.\  {\bf 89}, 171802 (2002)
                   \href{http://arXiv.org/abs/hep-ph/0205218}{[hep-ph/0205218]};
C.~Barbot and M.~Drees,
Phys.\ Lett.\  B {\bf 533}, 107 (2002) 
                   \href{http://arXiv.org/abs/hep-ph/0202072}{[hep-ph/0202072]};
C.~Barbot and M.~Drees,
Astropart.\ Phys.\  {\bf 20}, 5 (2003) 
                   \href{http://arXiv.org/abs/hep-ph/0211406}{[hep-ph/0211406]};
M.~Kachelriess and P.~D.~Serpico,
Phys.\ Rev.\  D {\bf 76}, 063516 (2007) 
                   \href{http://arXiv.org/abs/0707.0209}{[arXiv:0707.0209]};
N.~F.~Bell, J.~B.~Dent, T.~D.~Jacques and T.~J.~Weiler,
Phys.\ Rev.\  D {\bf 78}, 083540 (2008)
                   \href{http://arXiv.org/abs/0805.3423}{[arXiv:0805.3423]};
J.~B.~Dent, R.~J.~Scherrer and T.~J.~Weiler,
Phys.\ Rev.\  D {\bf 78}, 063509 (2008) 
                   \href{http://arXiv.org/abs/0806.0370}{[arXiv:0806.0370]};             
V.~Barger, Y.~Gao, W.~Y.~Keung, D.~Marfatia,
  Phys.\ Rev.\  {\bf D80}, 063537 (2009) 
                   \href{http://arXiv.org/abs/0906.3009}{[arXiv:0906.3009]};
  J.~F.~Fortin, J.~Shelton, S.~Thomas and Y.~Zhao,
                   \href{http://arXiv.org/abs/0908.2258}{arXiv:0908.2258};
M.~Kachelriess, P.~D.~Serpico and M.~A.~Solberg,
Phys.\ Rev.\  D {\bf 80}, 123533 (2009) 
                   \href{http://arXiv.org/abs/0911.0001}{[arXiv:0911.0001]}.

     
\bibitem{mdm}
  M.~Cirelli, N.~Fornengo and A.~Strumia,
  Nucl.\ Phys.\  B {\bf 753}, 178 (2006)
  	  \href{http://arXiv.org/abs/hep-ph/0512090}{[hep-ph/0512090]}.

\bibitem{splitting}
 J.~Hisano, S.~Matsumoto, M.~M.~Nojiri and O.~Saito,
  Phys.\ Rev.\  D {\bf 71}, 063528 (2005)
    	  \href{http://arXiv.org/abs/hep-ph/0412403}{[hep-ph/0412403]};
    H.~C.~Cheng, B.~A.~Dobrescu and K.~T.~Matchev,
  Nucl.\ Phys.\  B {\bf 543}, 47 (1999)
      	  \href{http://arXiv.org/abs/hep-ph/9811316]}{[hep-ph/9811316]}.

\bibitem{gluonrad}
  M.~Drees, G.~Jungman, M.~Kamionkowski, M.~M.~Nojiri,
  Phys.\ Rev.\  {\bf D49}, 636 (1994)
 \href{http://arXiv.org/abs/hep-ph/9306325}{[hep-ph/9306325]};
   M.~Asano, T.~Bringmann and C.~Weniger,
   \href{http://arXiv.org/abs/1112.5158}{[arXiv:1112.5158]}.


\bibitem{pythia}
T.~Sjostrand, S.~Mrenna, P.~Z.~Skands,
  JHEP {\bf 0605}, 026 (2006)
                   \href{http://arXiv.org/abs/hep-ph/0603175}{[hep-ph/0603175]};
T.~Sjostrand, S.~Mrenna, P.~Z.~Skands,
  Comput.\ Phys.\ Commun.\  {\bf 178}, 852 (2008)
                   \href{http://arXiv.org/abs/0710.3820}{[arXiv:0710.3820]}.

\bibitem{Bringmann:2011ye} 
  T.~Bringmann, F.~Calore, G.~Vertongen and C.~Weniger,
  Phys.\ Rev.\ D {\bf 84}, 103525 (2011)
                   \href{http://arXiv.org/abs/1106.1874}{[arXiv:1106.1874]}.
  
\bibitem{winomass}
  J.~Hisano, S.~Matsumoto, O.~Saito and M.~Senami,
  Phys.\ Rev.\ D {\bf 73}, 055004 (2006)
                   \href{http://arXiv.org/abs/hep-ph/0511118}{[hep-ph/0511118]};
    M.~Cirelli, A.~Strumia and M.~Tamburini,
  Nucl.\ Phys.\ B {\bf 787}, 152 (2007)
                 \href{http://arXiv.org/abs/0706.4071}{[arXiv:0706.4071]}.    

    
  \bibitem{futuro}
 A.~De Simone {\it et al.}, in progress.

\end{thebibliography}


\end{document}